\documentstyle[psfig,12pt]{article}
\newcommand{\ra}{\rangle}

\begin{document}

\begin{center}
{\bf {\Huge Quantum Reed-Muller Codes}}

\vspace{1cm}

Lin Zhang and Ian Fuss\\
Communications Division\\
Defence Science and Technology Organisation \\
P O Box 1500, Salisbury, South Australia 5108\\
Email: Lin.Zhang@dsto.defence.gov.au Ian.Fuss@dsto.defence.gov.au
\end{center}

\begin{abstract}
This paper presents a set of quantum Reed-Muller codes which are typically
100 times more effective than existing quantum Reed-Muller codes. The code
parameters are $[[n,k,d]]=[[2^m, \sum_{l=0}^{r}C(m,l)-\sum_{l=0}^{m-r-1}
C(m,l), 2^{m-r}]]$ where $2r+1>m>r$.
\end{abstract}

\subsection*{1. Introduction}

Quantum information processing, which includes quantum communication,
cryptography, and computation is currently moving from theoretical analysis
towards implementation \cite{spiller}. The fragility of quantum states
has led to the the development of quantum error correcting codes
\cite{shor1}\cite{steane1}. A large number of research papers have appeared
since it was shown that quantum error correcting codes exist
\cite{steane1}-\cite{9604034}. Much of this research work has focused on
repetition codes, i.e., codes with parameters [[n,1,d]], or single error
correcting codes, i.e., codes with parameters [[n,k,3]]. Repetition codes
are inefficient, with code rates ($R=\frac{1}{n}$). Single error correcting
codes are not very effective because they correct only one error in a block
of $n$ qubits. It is desirable to design quantum error correcting codes that
are more efficient, and more powerful.

Calderbank and Shor described a general method to construct non-trivial,
multiple error correcting quantum codes \cite{9512032}. The same method was
independently discovered by Steane \cite{9601029}. Using this method, together
with a technique to accommodate more information qubit(s) in a coded block
by slightly reducing the minimum Hamming distance, Steane presented a set of
multiple error correcting codes based on trivial codes and Hamming codes
\cite{9605021}, and a set of quantum Reed-Muller Codes based on classical
Reed-Muller codes \cite{9608026}. In this paper, a new set of quantum
Reed-Muller codes are derived using the technique of Calderbank
\& Shor/Steane. These codes have lower code rates than Steane's but
their minimum Hamming distances are larger. It will be shown that if the
uncoded quantum system has a qubit error rate less than $0.3\%$, one
particular code of rate $R=0.246$ will be able to bring the
output qubit error rate down to $10^{-9}$.

\subsection*{2. Quantum Codes}

A multiple error correcting quantum code described in \cite{9512032}
comprises two classical error correcting codes, $C_1=(n, k_1, d_1)$
and $C_2=(n, k_2, d_2)$. $C_1$ and $C_2$ are related by
\begin{equation}
\{0\}\subset C_2\subset C_1 \subset F_2^n
\label{eq1}
\end{equation}
where $\{0\}$ is the all-zero code word of length $n$,
and $F_2^n$ is the $n$-dimensional binary vector space. The quantum code
${\cal C}$ defined by $C_1$ and $C_2$ has the parameters
\begin{equation}
[[n,k,d]] = [[n,\dim(C_2^{\bot})-\dim(C_1^{\bot}), \min(d_1, d_2)]]
\label{eq2}
\end{equation}
where $C^{\bot}$ represents the dual code of $C$; the parameters of
$C^{\bot}$ are $(n, n-k, d^{\bot})$; $d$ and $d^{\bot}$ are indirectly
related by MacWilliams theorem \cite{MacWilliams}; and $\dim(C)$ denotes
the dimension of the code $C$.

In order for the error correction scheme to work, coded quantum states must
be represented in two bases. As in \cite{steane1}, we use ($|0\ra$, $|1\ra$)
to represent basis 1, and ($|{\bf 0}\ra = |0\ra+|1\ra$,
$|{\bf 1}\ra = |0\ra - |1\ra$) to represent basis 2. Note that the
normalisation coefficients have been omitted for the sake of simplicity in
presentation.

A quantum state, denoted by  $|w\ra$, can be coded as
\begin{equation}
|c_w\ra = \sum_{v\in C_1} (-1)^{vw}|v\ra \mbox{\hspace{0.75cm} in basis 1, and}
\label{eq3}
\end{equation}
\begin{equation}
|{{\bf c}}_w\ra = \sum_{v\in C_1^{\bot}}|v+w\ra \mbox{\hspace{2cm} in basis 2}
\label{eq4}
\end{equation}
where $w\in C_2^{\bot}/C_1^{\bot}$; $C_2^{\bot}/C_1^{\bot}$ denotes the
cosets of $C_1^{\bot}$ in $C_2^{\bot}$. Since all the $w's$ in
$C_1^{\bot}$ define the same quantum state, we can choose $w$ to
be coset leaders in $C_2^{\bot}/C_1^{\bot}$, denoted by
$w\in[C_2^{\bot}/C_1^{\bot}]$. There are
$|C_2^{\bot}/C_1^{\bot}| = 2^{\dim(C_2^{\bot})-\dim(C_1^{\bot})}$
coset leaders.  The quantum code ${\cal C}$ spans a $2^k$-dimensional Hilbert
space where $k = \dim(C_2^{\bot})-\dim(C_1^{\bot})$. It is
equivalent to encoding a block of $k$ qubits into $n$ qubits (mapping from
the $2^k$-dimensional Hilbert space into a subspace of a $2^n$-dimensional
Hilbert space).

\subsection*{3. Reed-Muller Codes}

Reed-Muller codes are self-dual and hence good candidates for
constructing quantum error correcting codes \cite{9608026}.
In this section, a brief introduction to the structure of Reed-Muller Codes
is given. A thorough treatment of this subject
can be found in \cite{MacWilliams}-\cite{Forney}.

For each positive integer $m$ and $r\ (0\leq r\leq m)$, there exists a
Reed-Muller code of block length $n=2^m$. This code, denoted by $RM(r,m)$,
is called the $r$-th order Reed-Muller code. The generator matrix of
$RM(r,m)$ is defined as
\begin{equation}
{\bf G} = \left[\begin{array}{c}
	{\bf G}_0\\ {\bf G}_1 \\ \vdots \\ {\bf G}_r \end{array}\right]
\label{eq5}
\end{equation}
where ${\bf G}_0=\{1\}$ is the all-one vector of length $n$; ${\bf G}_1$,
an $m$ by $2^m$ matrix, has each binary $m$-tuple appearing once as a column;
and ${\bf G}_l$ is constructed from ${\bf G}_1$ by taking its rows to be all
possible products of rows of ${\bf G}_1$, $l$ rows of ${\bf G}_1$ to a product.
For definiteness, we take the leftmost column of ${\bf G}_1$ to be all zeros,
the rightmost to be all ones, and the others to be the binary $m$-tuples
in increasing order, with the low-order bit in the bottom row.
Because there are $C(m,l)$ ways to choose the $l$ rows in a product,
${\bf G}_l$ is a $C(m,l)$ by $2^m$ matrix. For an $r$-th order Reed-Muller
code, the dimension of the code is given by $k=\sum_{l=0}^{r}C(m, l)$.

Equation (\ref{eq5}) shows that $RM(r-1,m)$ is generated by
\([{\bf G}_0, {\bf G}_1, \cdots, {\bf G}_{r-1}]^T\), therefore
$RM(r-1,m)\subset RM(r,m)$. More generally,
\begin{equation}
RM(r-i,m)\subset RM(r,m) \mbox{\hspace{2cm} for integers \hspace{1cm}}
1\leq i\leq r.
\label{subset}
\end{equation}

The self-duality property and the minimum Hamming distance of a Reed-Muller
code can be easily derived from the squaring structure of the code.
According to Forney \cite{Forney}, any $r$-th order Reed-Muller code
of length $2^m$ can be generated through recursive squaring construction:
\begin{equation}
RM(r,m) = |RM(r,m-1)/RM(r-1,m-1)|^2
\label{sq1}
\end{equation}
or two-level squaring construction:
\begin{equation}
RM(r,m) = |RM(r,m-2)/RM(r-1,m-2)/RM(r-2,m-2)|^4
\label{sq2}
\end{equation}
where $RM(r,m-1)/RM(r-1,m-1)$ and $RM(r,m-2)/RM(r-1,m-2)/RM(r-2,m-2)$
represent a one-level partition and a two-level partition, respectively.
The boundary conditions are:
\[
RM(r,0)=\left\{\begin{array}{ll}(1,1) & \mbox{if $r\geq 0$} \\ (1,0) &
 \mbox{otherwise} \end{array} \right.
\]
The squaring construction of $RM(r,m)$ is defined as
\begin{eqnarray}
\lefteqn{|RM(r,m-1)/RM(r-1,m-1)|^2} \nonumber \\
 &=&\{(t_1+c,t_2+c): \nonumber \\
 & &t_1,t_2\in RM(r-1,m-1), c\in
[RM(r,m-1)/RM(r-1,m-1)]\}.
\label{sq3}
\end{eqnarray}
From this construction, it is obvious that the minimum Hamming distance
of $RM(r,m)$, denoted $d[RM(r,m)]$, is given by
\begin{equation}
d[RM(r,m)]=\min\{d[RM(r-1,m-1)], 2d[RM(r,m-1)]\}.
\label{sq4}
\end{equation}
From the boundary condition, one can easily prove by induction that the
minimum Hamming distance of $RM(r,m)$ is indeed $2^{m-r}$.

The dual partition chain of $RM(r,m-1)/RM(r-1,m-1)$ is
$RM^{\bot}(r-1,m-1)/RM^{\bot}(r,m-1)$. The squaring construction of this
dual partition chain is written as
\begin{eqnarray}
\lefteqn{|RM^{\bot}(r-1,m-1)/RM^{\bot}(r,m-1)|^2}  \nonumber \\
 &=&\{(t'_1+c',t'_2+c'): \nonumber \\
 & &t'_1,t'_2\in RM^{\bot}(r,m-1),c'\in [RM^{\bot}(r-1,m-1)/RM^{\bot}(r,m-1)]\}.
\label{sq5}
\end{eqnarray}
$|RM^{\bot}(r-1,m-1)/RM^{\bot}(r,m-1)|^2$ is the dual of
$|RM(r,m-1)/RM(r-1,m-1)|^2$ because the inner product of the vectors from
the two constructions is zero,
\begin{equation}
(t_1+c,t_2+c)\cdot(t'_1+c',t'_2+c')=(c,c)\cdot(c',c')=0.
\label{sq6}
\end{equation}
If $RM^{\bot}(r,m-1)=RM[(m-1)-r-1,m-1]$ for $r\leq m-1$,
Eqs (\ref{sq5}) and (\ref{sq6}) shows that $RM^{\bot}(r,m)=RM(m-r-1,m)$
for $r\leq m$. Using the induction method and the boundary condition for
the initial partition, it can be shown that the dual code of
$RM(r,m)$ is $RM(m-r-1,m)$. If $m-r-1\leq r$,
\[
RM^{\bot}\equiv RM(m-r-1,m)\subseteq RM(r,m).\]
That is, Reed-Muller codes are self-dual.
The minimum Hamming distance of the dual code
$RM(m-r-1,m)$ is $2^{r+1}$.
Reed-Muller Codes of block length $4$ to $1024$ are listed in
Table~\ref{table1}.

\subsection*{4. Construction}

Since Reed-Muller codes are self-dual, we have $C^{\perp}\subseteq C$
if $k^{\perp}\leq k$. Let $C_1\equiv (n,k,d)\equiv (2^m, \sum_{l=0}^{r}C(m,l),
2^{m-r})$ and $k\geq \frac{n}{2}$.
To construct a quantum code from Reed-Muller codes,
we simply choose $C_2=C_1^{\perp}$. Therefore,
$C_2\equiv (n,k^{\perp},d^{\perp})\equiv (2^m, \sum_{l=0}^{m-r-1}C(m,l), 
2^{r+1})$ and $d^{\perp}\geq d$.
Applying the method introduced in Section 2 and Equations
(\ref{eq2})-(\ref{eq4}), we will have the
quantum Reed-Muller code ${\cal C}$ defined as
\begin{equation}
{\cal C}=[[n,k,d]]=[[2^m, \sum_{l=0}^{r}C(m,l)-\sum_{l=0}^{m-r-1}C(m,l),
2^{m-r}]]
\end{equation}
For example, let $m=10$ and $n=2^m=1024$. There are 10 classical
Reed-Muller codes of length 1024:
\[\begin{array}{ccccc}
(1024,1023,2)&(1024,1013,4)&(1024,968,8)&(1024,848,16)&(1024,638,32) \\
(1024,1,1024)&(1024,11,512)&(1024,56,256)&(1024,176,128)&(1024,386,64)\\
\end{array}
\]
Every pair of dual codes form a quantum Reed-Muller code:
\[\begin{array}{ccccc}
(1024,1022,2)&(1024,1002,4)&(1024,912,8)&(1024,772,16)&(1024,252,32)
\end{array}
\]
The $(1024,252,32)$ code can correct $15$ errors out of $1024$ qubits.
A comparable code in \cite{9608026} is the $(1024,462,24)$ code, which
is about $1.8$ times more efficient than the $(1024,252,32)$ code. But
the $(1024,252,32)$ code is able to correct 4 more errors than the
$(1024,462,24)$ code.

A list of new quantum Reed-Muller codes of length $4$ to $1024$ is given in
Table~\ref{table2}. These codes, together with codes listed in
\cite{9608026}, form one family of quantum Reed-Muller codes.

Assume that the decoherence process affects each qubit independently, and
that the error probability of uncoded qubits is $p$. Then the probability
of each coded quantum state in error shall be bounded by
\begin{equation}
P_e\leq\sum_{j=t+1}^{n}C(n, j)p^{n-j} (1-p)^j.
\label{bound1}
\end{equation}
The probability of each qubit being in error is given by
\begin{equation}
P_q = 1-(1-P_e)^{\frac{1}{n}}
\label{bound2}
\end{equation}

\begin{figure}[hbtp]
\begin{center}
\ \psfig{file=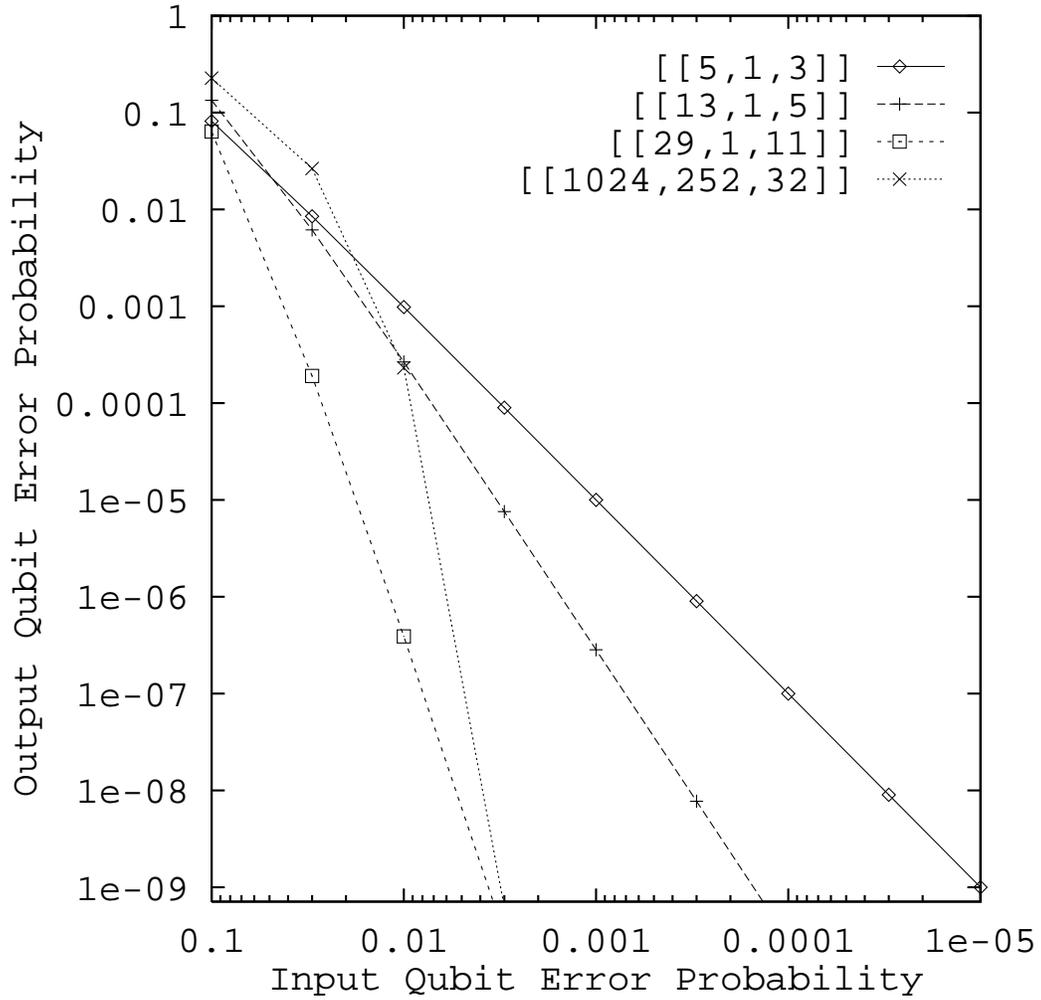,width=14cm}
\end{center}
\caption{Qubit Error Performance of repetition codes and the (1024,252,32) Reed-Muller code.}
\label{qber1}
\end{figure}

\begin{figure}[hbtp]
\begin{center}
\ \psfig{file=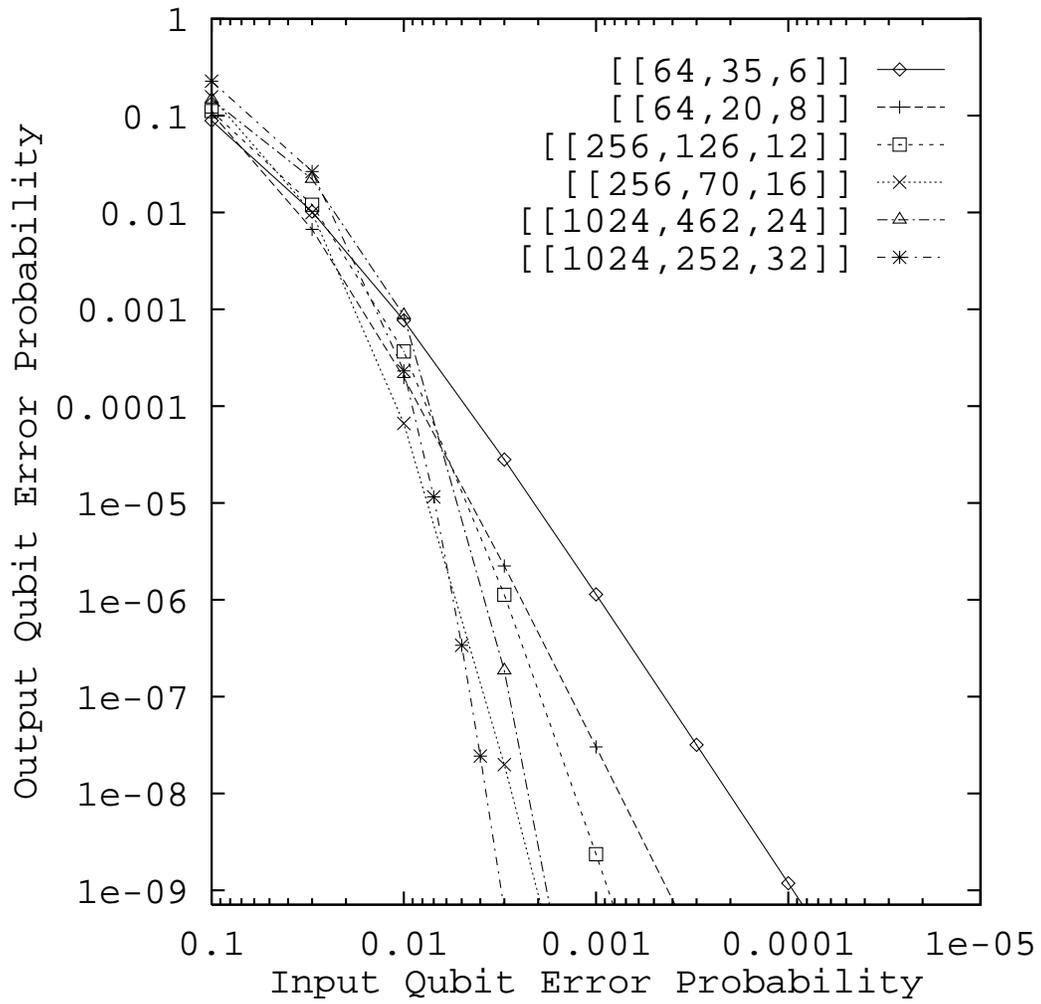,width=14cm}
\end{center}
\caption{Qubit Error Performance of Multiple Error Correcting Codes.}
\label{qber2}
\end{figure}

The error performances of various quantum error correcting codes are
illustrated in Figs.\ref{qber1} and \ref{qber2} by applying
Equations~(\ref{bound1}) and (\ref{bound2}). It was found that
all single error correcting quantum codes have similar qubit error
performance (close to that of the $[[5,1,3]]$ code). The most effective
repetition codes were proposed by Calderbank et. al. \cite{9605005}.
Fig.\ref{qber1} shows that the $[[1024,252,32]]$ quantum Reed-Muller
code outperforms the $[[13,1,5]]$ repetition code significantly, and
outperforms the $[[29,1,11]]$ repetition code asymptotically. The
$[[1024,252,32]]$ code is about $3$ times more efficient than the
$[[13,1,5]]$ code, and $7$ times more efficient than the
$[[13,1,5]]$ code. It should be noted though that codes such as
$[[1024,252,32]]$ are more complex to decode. 

The quantum Reed-Muller codes constructed in this paper are a factor
of $2$ less efficient than those constructed by Steane \cite{9608026}.
However they are two orders of magnitude more effective, as shown
in Fig.\ref{qber2}. As indicated in both figures, the average qubit
error probability can be reduced to less than $10^{-9}$ (one
in a billion) if the uncoded qubit error rate is not more than $0.3\%$.

\clearpage
\pagebreak

\begin{table}
\begin{center}
\begin{tabular}{lr|rrrrrrrrrr}
k &      & \multicolumn{10}{c}d \\
  &      &    2 &    4 &    8 &    16 &  32 &  64 & 128 & 256& 512& 1024\\
									\hline
  &    4 &    3 &    1 &      &       &     &     &     &    &    &   \\
  &    8 &    7 &    4 &    1 &       &     &     &     &    &    &   \\
  &   16 &   15 &   11 &    5 &     1 &     &     &     &    &    &   \\
n &   32 &   31 &   26 &   16 &     6 &   1 &     &     &    &    &   \\
  &   64 &   63 &   57 &   42 &    22 &   7 &   1 &     &    &    &   \\
  &  128 &  127 &  120 &   99 &    64 &  29 &   8 &   1 &    &    &   \\
  &  256 &  255 &  247 &  219 &   163 &  93 &  37 &   9 &  1 &    &   \\
  &  512 &  511 &  502 &  466 &   382 & 256 & 130 &  46 & 10 &  1 &   \\
  & 1024 & 1023 & 1013 &  968 &   848 & 638 & 386 & 176 & 56 & 11 & 1 \\
\end{tabular}
\end{center}
\caption{Parameters of the Classical Reed-Muller Codes}
\label{table1}
\end{table}

\begin{table}
\begin{center}
\begin{tabular}{lr|rrrrr}
k &      & \multicolumn{5}{c}d \\
  &      &    2 &    4 &    8 &    16 &    32 \\ \hline
  &    4 &    2 &      &      &       &       \\
  &    8 &    6 &    0 &      &       &       \\
  &   16 &   14 &    6 &      &       &       \\
n &   32 &   30 &   20 &    0 &       &       \\
  &   64 &   62 &   50 &   20 &       &       \\
  &  128 &  126 &  118 &   68 &     0 &       \\
  &  256 &  254 &  238 &  184 &    70 &       \\
  &  512 &  510 &  492 &  420 &   252 &     0 \\
  & 1024 & 1022 & 1002 &  912 &   772 &   252 \\
\end{tabular}
\end{center}
\caption{Parameters of the Quantum Reed-Muller Codes}
\label{table2}
\end{table}

\end{document}